\documentclass[conference]{IEEEtran}
\usepackage{graphicx,times,amsmath, amssymb}
\usepackage{fixltx2e} 
\usepackage{xspace}
\usepackage{url}
\usepackage{a4wide}
\usepackage{paralist}
\usepackage[caption=false,font=footnotesize]{subfig}
\usepackage{epstopdf}
\usepackage{graphicx}

\newcommand{\ie}{i.e.\xspace}

\begin{document}

\title{MATCASC: A tool to analyse cascading line outages in power grids}

\author{Yakup Ko\c{c}$^{1}$~~Trivik Verma$^2$~~Nuno A. M. Araujo$^2$~~Martijn Warnier$^1$\\
$^1$Faculty of Technology, Policy and Management, Delft University of Technology\\
$^2$ Computational Physics for Engineering Materials, ETH Zurich
}
\date{}

\maketitle
\thispagestyle{empty}

\begin{abstract}
Blackouts in power grids typically result from cascading failures. The key importance of the electric power grid to society encourages further research into sustaining power system reliability and developing new methods to manage the risks of cascading blackouts. Adequate software tools are required to better analyse, understand, and assess the consequences of the cascading failures. This paper presents MATCASC, an open source MATLAB based tool to analyse cascading failures in power grids. Cascading effects due to line overload outages are considered. The applicability of the MATCASC tool is demonstrated by assessing the robustness of IEEE test systems and real-world power grids with respect to cascading failures.

\end{abstract}

\section{Introduction}
\label{sec_Introduction}
The electric power grid is essential for modern society. The continuous availability and reliability of the power grid is crucial for the daily life routines of millions of people. However, large-scale blackouts repeatedly occur~\cite{BlackoutRefBrazil,blackoutReport}. 
Blackouts have a direct cost of tens of billions of dollars annually~\cite{Baldick2008}. 
Due to the strong interdependencies, failures might also propagate into other critical infrastructures such as telecommunications, transportation, and water supply~\cite{Eeten2011, Schneider2013, Buldyrev2010}. The key importance of the electric power grid to society encourages further research into sustaining power system reliability and developing new methods to evaluate and mitigate the risks of cascading blackouts. Adequate software tools are required to better analyse, understand, and assess the consequences of the cascading failures in power grids.


Cascading failure is a sequence of dependent failures of individual components that successively weakens the power system~\cite{Baldick2008}. The cascade of an initial failure occurs in different ways, including the instability of voltage and frequency, hidden failures of protection systems, software, or operator error, and line overloads. Different modelling studies focus on a small subset of these mechanisms. This is a necessary and healthy approach at this stage of the development of the field~\cite{Vaiman2012}. The authors in \cite{Kirschen2004,Hardiman2004} propose models having approximate representations of protection, operator actions and voltage collapse by deploying AC load flow analysis, while the researchers in~\cite{Carreras2004, Chen2005, Liao2004} present models of cascading failures due to line overloads deploying a DC load flow analysis. Carreras et al.~\cite{Carreras2004} represent the slow complex dynamics of grid upgrade while Chen et al.~\cite{Chen2005} model hidden failures of the protection system. Although different modelling perspectives have already been proposed, no effort is devoted to make a non-commercial tool available for researchers to work with. This paper presents MATCASC, an open source MATLAB based package to analyse cascading failures due to line overloads in a power grid. MATCASC simulates cascading line outages rather than controlling and mitigating them. The MATCASC tool is mainly intended to be used for academic purposes, and to be extended to incorporate also other aspects of cascading failures in power grids. The open source MATCASC tool provides a basis towards a more comprehensive cascading failure analysis tool.

Section~\ref{sec_Modelling Cascading Failures in Power Grids} models the concepts/components of the cascading failures in power grids. Section~\ref{sec_Architecture of MATCASC} introduces the architecture of the MATCASC and elaborates on the main modules. Section~\ref{sec_Use Cases} demonstrates the applicability of MATCASC via use cases while Section~\ref{sec_Conclusion} ends the paper with a conclusions and suggestions for the future work. 

%
%

\section{Modelling Cascading Failures in Power Grids}
\label{sec_Modelling Cascading Failures in Power Grids}

This section develops a model to simulate cascading failures due to line overloads in power grids by using the complex network approach. This requires modelling a power grid as a graph, estimating line flows across the grid, and maximum line capacities, and modelling line protections.

\subsection{Mapping the power grid topology to a graph}
\label{subsec_Mapping the power grid topology to a graph}
A power grid is a multiplex network that is composed of three functional parts: generation, transmission, and distribution. Power is provided from generation buses to distribution stations through the transmission buses that are all inter-connected via transmission lines. In a graph representation of a power grid, nodes represent generation, transmission, distribution buses, substations and transformers, while links model the transmission lines. All the parallel transmission lines in a single-line diagram are represented by an equivalent single link in a graph representation. Additionally, the links in a graph representation are weighted by the admittance (or impedance) value of the corresponding transmission line.
%

\subsection{Estimating line power flows: DC load flow equations}
\label{subsec_Estimating line power flows: DC load flow equations}

Electric power flows in a grid according to Kirchoff laws. Accordingly, the physical properties of a grid including impedances, voltage levels at each individual power station, voltage phase differences between power stations and loads at terminal stations control the power flow in the grid. The power flow equations are the basic tool to estimate the flow values for each component in the network. AC power flow equations are non-linear equations that model the flows of both active and reactive powers. DC load flow equations are a simplified and linearised version of AC power flow equations considering only flow of active power~\cite{Grainger1994, Hertem2006}. This model deploys DC load flow analysis to estimate the flow values across the network rather than AC load flow equations because AC load flow analysis may not converge when the lines and generators trip and introduces significant complexity in the model~\cite{Bao2009}. In the AC model, the active power flow \emph{$f_{ij}$} through a transmission line \emph{$l_{ij}$} connecting node \emph{i} and node \emph{j} is related to the complex voltage at both nodes \emph{i} and \emph{j} and the impedance of the line \emph{$l_{ij}$} as follows~\cite{Hertem2006}:

\begin{equation}\label{AC flow}
f_{ij}=\frac{|V_{i}| |V_{j}|}{z_{ij}}\sin(\theta_{ij})
\end{equation}

\noindent where \emph{$|V_{i}|$} is the voltage amplitude at node \emph{i}, \emph{$\theta_{ij}$} is the voltage phase difference between node \emph{i} and node \emph{j}, and \emph{$z_{ij}$} is the impedance of transmission line \emph{$l_{ij}$}. The above non-linear equation is linearised by the following assumptions:

\begin{itemize}
\item line resistance is ignored so that line impedance equals line reactance \emph{$z_{ij}$} $\approx$ \emph{$x_{ij}$}; 
\item voltage phase differences are very small so that \emph{$\sin(\theta_{ij})$} $\approx$ $\theta_{ij}$;
\item flat voltage profile.
\end{itemize}

By incorporating the above assumptions in Eq.~\eqref{AC flow}, the DC power flow equation is obtained:

\begin{equation}\label{DC flow}
f_{ij}=\frac{\theta_{ij}}{x_{ij}}=b_{ij}\theta_{ij}
\end{equation}

\noindent where \emph{$b_{ij}$} is the susceptance of line \emph{$l_{ij}$}. The entire system can be modelled solely by using the following linear equation:

\begin{equation}\label{DC flow summation}
P_{i}=\sum_{j=1}^{d} f_{ij}=\sum_{j=1}^{d} b_{ij} \theta_{ij}
\end{equation}

\noindent where \emph{$P_{i}$} is the real power flow at node \emph{i} and \emph{d} is the degree of node \emph{i}. In terms of matrices, Eq.~\eqref{DC flow summation} can be rewritten as:
\begin{equation}\label{DC flow equations}
\textbf{P}=\textbf{B}\boldsymbol\theta
\end{equation}

\noindent where \textbf{P} is the vector of real power injections, $\boldsymbol\theta$ contains the voltage angles at each node, and \textbf{B} is the bus susceptance matrix in which \emph{$B_{ij}=-\frac{1}{x_{ij}}$} and \emph{$B_{ii}=\sum_{j=1}^{d}-B_{ij}$}. Since the losses in the system are neglected, all the active power injections are known in advance. Hence, given the bus susceptance matrix \textbf{B}, the voltage angles at each node can be calculated directly by using:

\begin{equation}
\label{DC flow equations_phases}
\boldsymbol\theta=\textbf{B}^{\textbf{-1}}\textbf{P}
\end{equation}

After obtaining the voltage angle values at each node, the power flow values through each line can be computed by using Eq.~\eqref{DC flow}.

\subsection{Estimating line capacities}
\label{subsec_Estimating line capacities}
The maximum capacity of a line is defined as the maximum power flow that can be afforded by the line. The flow limit of a transmission line is imposed by thermal, stability or voltage drop constraints~\cite{Glover2001}. In real world infrastructures, the maximum capacity of a transmission line is constrained by cost~\cite{Motter2002}. This model assumes that the maximum capacity of a line relates to its initial load as follows: 

\begin{equation}
\label{eq:capEstimation}
C_{i}=\alpha_{i} L_{i}(0)
\end{equation}

where \emph{$C_{i}$} is the maximum capacity, \emph{$L_{i}(0)$} is the initial load, and $\alpha_{i}$ is the tolerance parameter of line \emph{$i$}.

\subsection{Modelling line protection}
\label{subsec_Modeling line protection}
In a power grid, transmission lines are protected by circuit breakers. A circuit breaker trips a line before the load of the line exceeds a certain threshold level in order to prevent that the transmission line is permanently damaged due to overloading. For the sake of simplicity, this paper assumes a deterministic model for line tripping mechanism. A circuit breaker for line $l$ trips at the moment the load of the line $l$ exceeds its maximum capacity: $|\textit{loading level}|>1$.

This model does not consider the malfunctioning of the protection devices also known as \textit{hidden failures}~\cite{Chen2005}, in order not to confound the model with details. While hidden failures contribute to cascading failures in power grids, they can be considered as having a second order impact to the impact of the interconnection of topology~\cite{Pepyne2007}. 

\section{The Architecture of MATCASC}
\label{sec_Architecture of MATCASC}

\begin{figure}[!htb]
\centering
\includegraphics[scale=0.37]{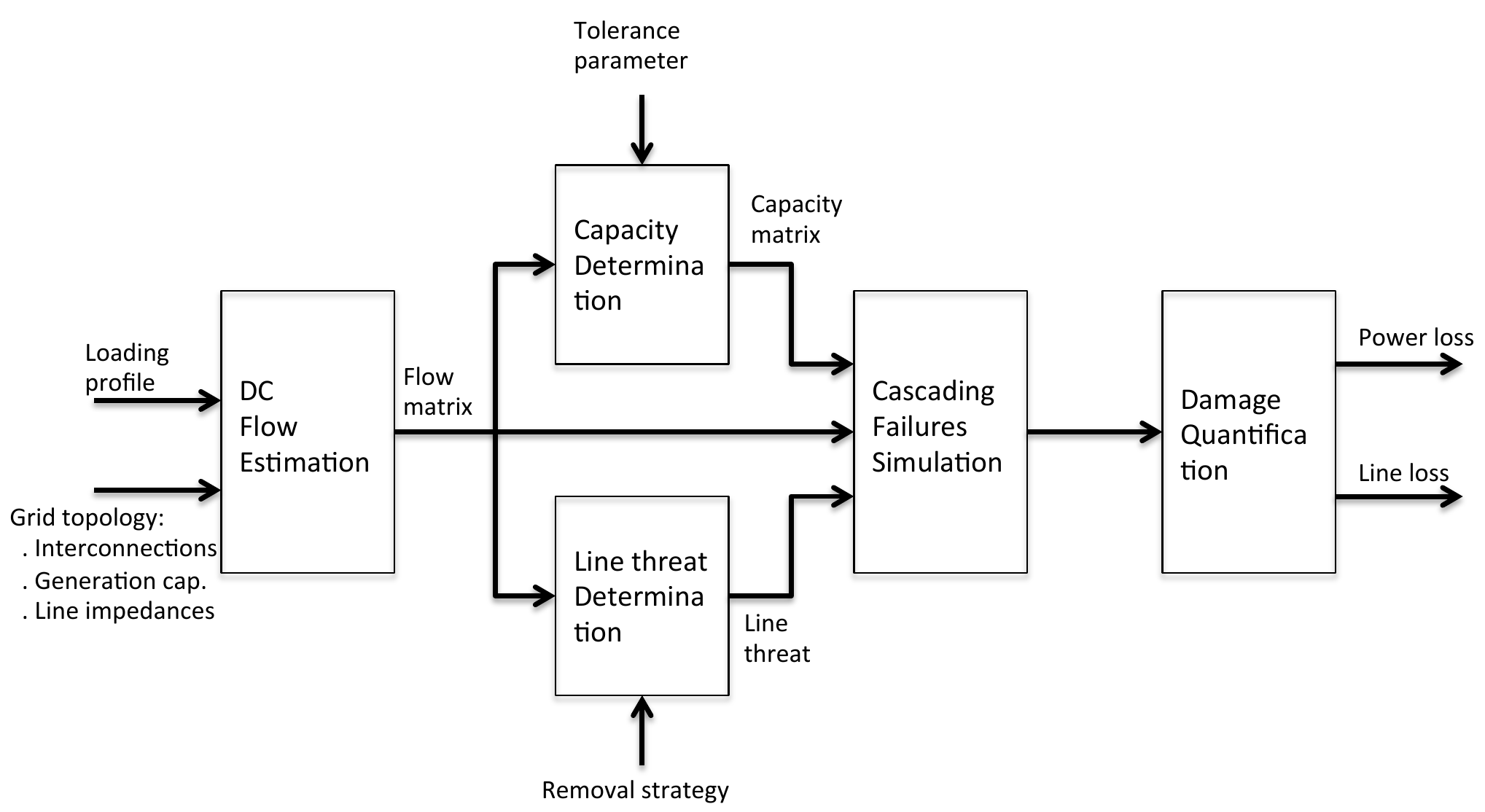}
\caption{MATCASC's architecture}
\label{fig:blackBox}
\end{figure}

MATCASC provides a flexible environment to simulate cascading failures due to line overloads in electric power grids and to analyse the damage after cascading failures occur. Its architecture, presented in Figure~\ref{fig:blackBox} with input-output relationships, is built from several core modules: (a) estimating load flow values across the grid, (b) estimating the capacity matrix for the given configuration, (c) determining the line threat, (d) simulating the cascading effect in the grid and (e) quantifying the damage in the grid due to the cascades.
 
 \subsection{Input}
\label{sec_Input} 
In order to analyse the cascading failures in a power grid, MATCASC requires (\textit{i}) a grid structure including the interconnection of the components with their specific electrical properties, (\textit{ii}) a loading profile for the grid, (\textit{iii}) a tolerance parameter for transmission lines, and (\textit{iv}) a removal strategy to determine the line threat in the grid. (\textit{i}) and (\textit{ii}) are provided as an input casedata file~\cite{Zimmerman2011} while (\textit{iii}) and (\textit{iv}) are directly given as input to the corresponding modules. 

\subsection{DC line power flow and capacity estimation }
\label{sec_DC line power flow estimation}

The first step to analyse a cascading failure in a power grid configuration to dispatch the power over the grid. The DC Flow Estimation module performs a DC load flow analysis to estimate line flow values based on the modelling perspective given in Sec.~\ref{subsec_Estimating line capacities}. The output of this module is a matrix including the flow values in the lines. An element \textit{$f_{ij}$} in the flow matrix indicates the active power flow value from bus \textit{i} to bus \textit{j}. Having obtained the flow values through each line, the Capacity Determination module computes the line capacity values by using the tolerance parameter input and Eq.(\ref{eq:capEstimation}). 

\subsection{Line threat determination}
\label{sec_Line threat determination}

The Line threat Determination module determines a line-to-be-removed to initiate a cascading failures in the power grid depending on the removal strategy input. The Line threat is the output. It is a row vector in the form of $[i  j]$ indicating the line between buses \textit{i} and \textit{j}. A removal strategy can be either based on a random removal or based on a targeted attack. MATCASC posses two different attack options: (i) an edge-betweenness centrality and (ii) an electrical node significance based attack.

The edge betweenness centrality~\cite{Mieghem2006} of a link is a graph topological metric quantifying the topological centrality of a link in a network. Edge betweenness centrality of a link \textit{l} is defined as the normalized number of shortest paths between any pair of nodes passing through \textit{l}.
%
%
%

The electrical node significance~\cite{Koc2013, Koc2013_2} is a contextual node centrality measure, specifically designed for power grids. It is defined as the amount of the outgoing power normalized by the total amount of the distributed power in the grid:

\begin{equation}\label{Delta}
\delta _{i}=\frac{P_{i}}{\sum_{j=1}^{N} P_{j}}
\end{equation}

\noindent  where $P_{i}$ stands for total power distributed by node $i$ while $N$ refers to number of nodes in the network.
 
 In an edge-betweenness centrality based removal strategy, the line with the highest betweenness is marked as the line threat while in an electrical node significance based attack, the most important link of the node with the highest electrical node significance is determined as the line threat.   
 
 \subsection{Cascading line outages simulation}
\label{sec_Cascading line outages simulation}

To simulate a cascading failure, the line-to-be-removed is pruned from the topology. MATCASC implements the removal of a line by setting the admittance value of the corresponding line to zero (\ie impedance value to infinity) so that the line flow is set to zero. This method models the effect of a line outage appropriately and avoids the singularity of the admittance matrix that would result from removing the outage. 
As a result of removal of overloaded links, some parts of the grid may disintegrate into separate islands. Disintegration of the grid is checked at every cascading stage whenever the overloaded lines are removed. As a measure to assess the grid connectivity, the rank of admittance matrix \textbf{B} is used. A full rank \textbf{B} indicates a connected grid while a grid disintegration is implied when \textbf{B} loses its rank.

\begin{figure}[!htb]
\centering
\includegraphics[scale=0.40]{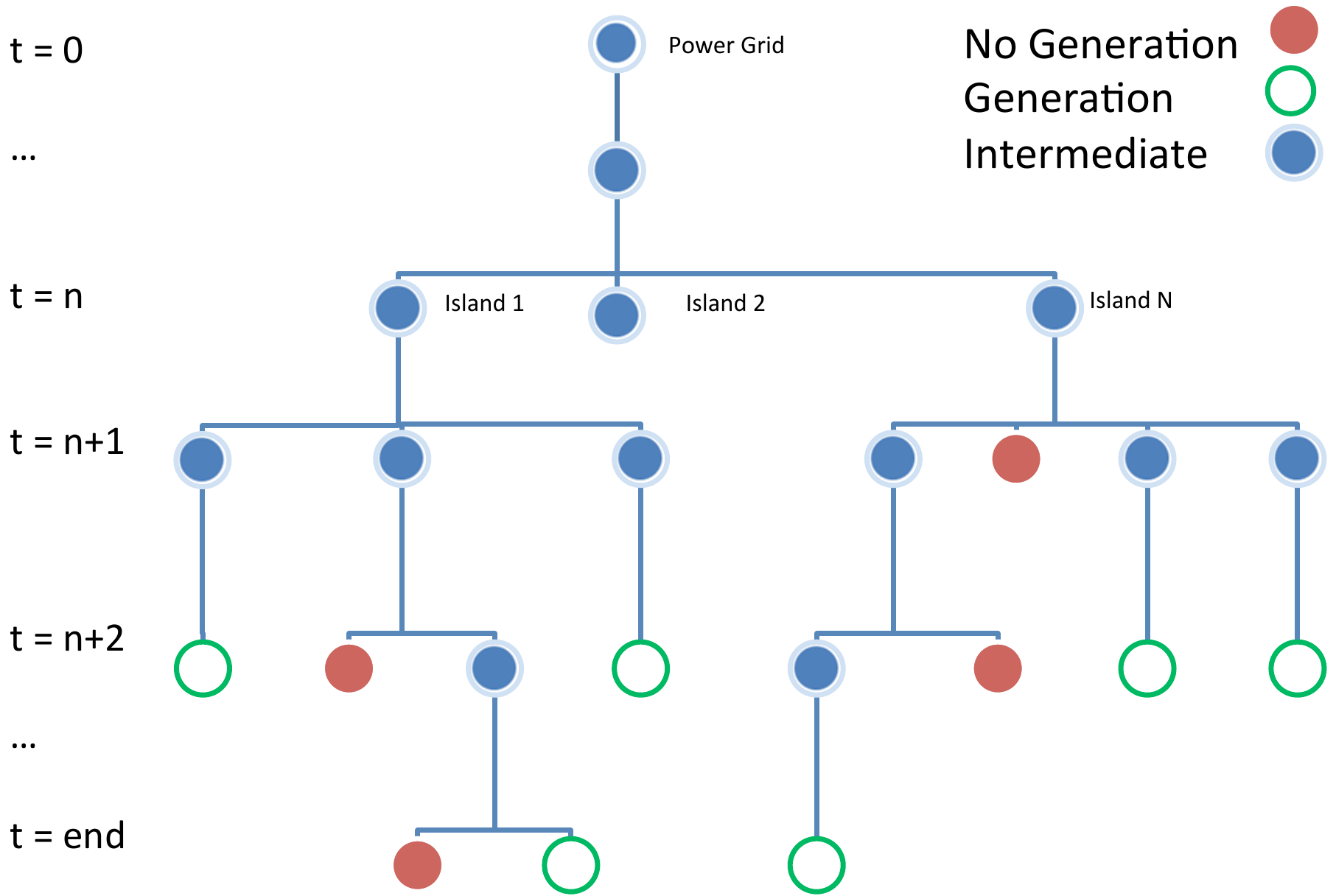}
\caption{A tree structure depicting the logical depth of the model and iterating procedure for islanding process}
\label{fig:Tree}
\end{figure}

Each island of the disintegrated network may or may not have generation substations. If an island is free of any generating source then it is dead, marked red in Fig.~\ref{fig:Tree} that depicts the iterating procedure for the islanding process due to the cascading effect per cascading stage $t$. Some islands may not be affected completely by flow redistribution, i.e. they still have a power generating source with enough capacity. These islands still satisfy their demand and are marked green in Fig.~\ref{fig:Tree}. All the intermediate islands in which cascading failures continues are marked blue. The tree structure shows how an island can undergo changes in the coarse of cascade and disintegrate into several islands some of which are functioning and some are deeply affected by the cascade.

The Cascading Failure Simulation module still iterates on all intermediate (blue) islands. The electric power is redistributed. The redistribution of the power flow may cause further successive line overloads and removals in the grid. Consequently, each intermediate island may result in an additional group of islands and the model iterates on each group of islands belonging to the same cascading stage simultaneously. Fig.~\ref{fig:Tree} illustrates the tree structure encapsulating this procedure.

\subsection{Quantifying the damage by cascading failures}
\label{subsec_Quantifying the damage by cascading effect}

After a cascading failure takes place in a power grid, the damage in the grid is quantified empirically by the metrics \emph{Power loss} and \emph{Link loss}. Power loss is the fraction of the not-satisfied power demand, while Link loss is defined as the fraction of de-energized lines after a cascading failure occurs in an electric power grid. A line is considered to be de-energized if it is tripped by its protection mechanism or if it is disconnected and isolated from generators (e.g. remains in a dead island) after the cascading failure. 
%
%
%

\section{Use Cases }
\label{sec_Use Cases}
%

This section evaluates the robustness of different test systems to exemplify for which purposes MATCASC can be used. To demonstrate the applicability of MATCASC, (i) the robustness of power grids with respect to cascading failures is assessed under targeted attacks, and (ii) the effectiveness of different attack strategies is investigated. 

\subsection{Robustness assessment of power grids under targeted attacks}
\label{subsec_Robustness assessment of different power grids under targeted attacks}

For a cascading failure robustness analysis, MATCASC requires the input data given in Sec.~\ref{sec_Input}. The IEEE test systems and the real-world Union for the Coordination of Transport of Electricity (UCTE) network provides the required information for a robustness analysis. Therefore, IEEE 57, IEEE 118, and the UCTE network with a summer loading profile are considered. For each network, a tolerance parameter of 1.5 (i.e. a loading level of 66\%) is chosen. In these grids, a cascading failure is triggered by targeted attacks. An attack strategy based on the electrical node significance metric is deployed. A targeted attack on these grids results in a cascading failure, which relates directly to the robustness of the grid. 

%
\begin{figure}[!htb]
\centering
\includegraphics[scale=0.2]{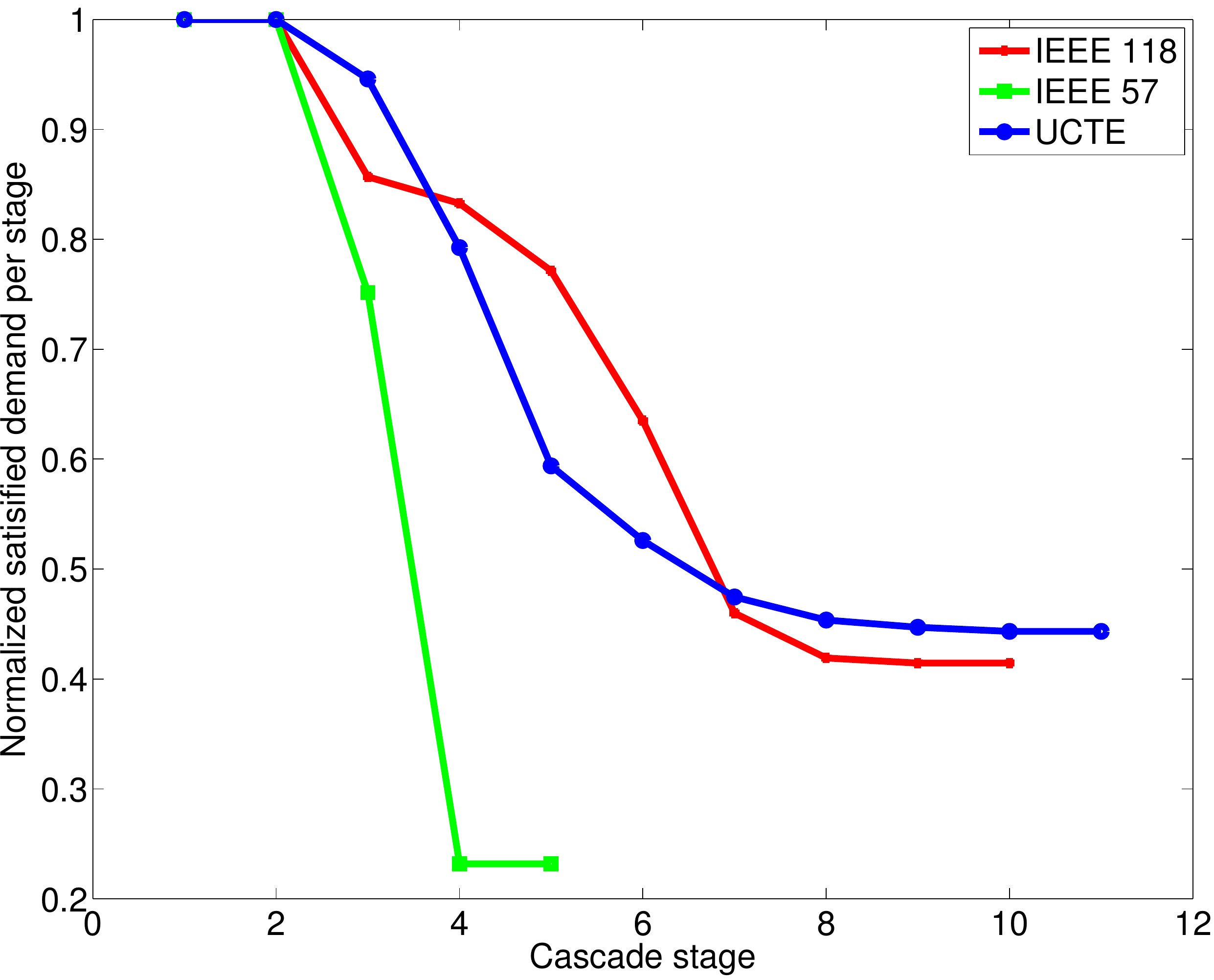}
\caption{Normalized satisfied power demand per stage of the cascade in different power grids}
\label{fig:robustnessCurves}
\end{figure}

The simulation results on the robustness assessment of the test configurations in Fig.~\ref{fig:robustnessCurves} suggests that the real world UCTE network is the most robust configuration against cascading failures due to line overload under targeted attacks while the IEEE 57 test system is the least robust configuration. An attack on the IEEE 57 system based on the electrical node significance results in a loss of almost 75\% of the total power demand. Additionally, the initial failure propagates into the grid very quickly, i.e. a small cascading depth. On the other hand, in the UCTE case, an attack causes that almost 50\% of the total power demand can not be served with a larger cascading depth. A larger cascading depth gives more time to the network operator to take preventive actions to better manage the cascading failures.

\subsection{Effectiveness assessment of cascade-induced attack strategies in power grids}
\label{sec_Effectiveness of cascade-induced attack strategies in power grids}
\begin{figure*}[!htb]
\begin{center}	
	\subfloat[IEEE 118]{
	\includegraphics[width=.35\textwidth]{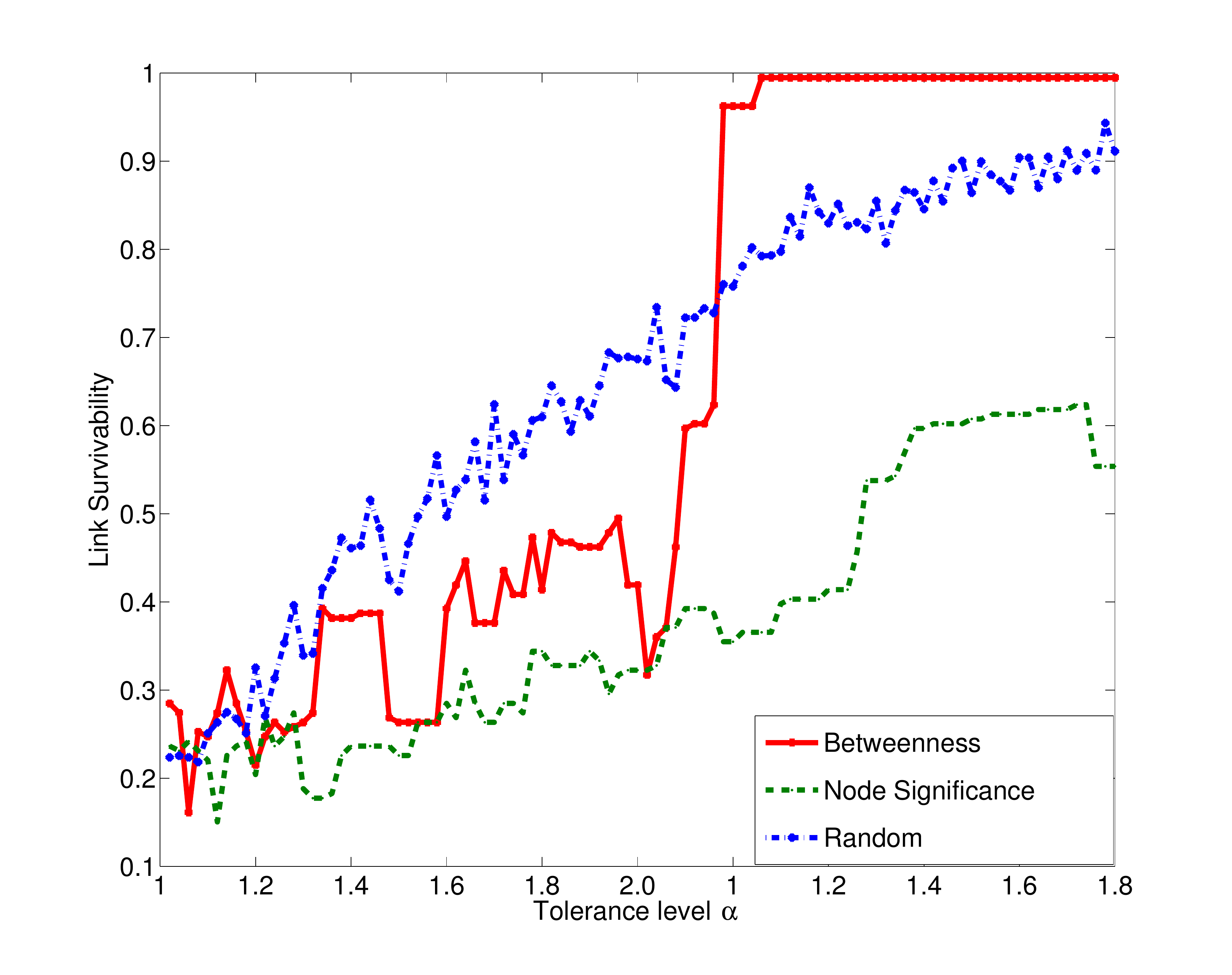}
	\label{fig:IEEELS}
	}
	\subfloat[UCTE]{
	\includegraphics[width=.35\textwidth]{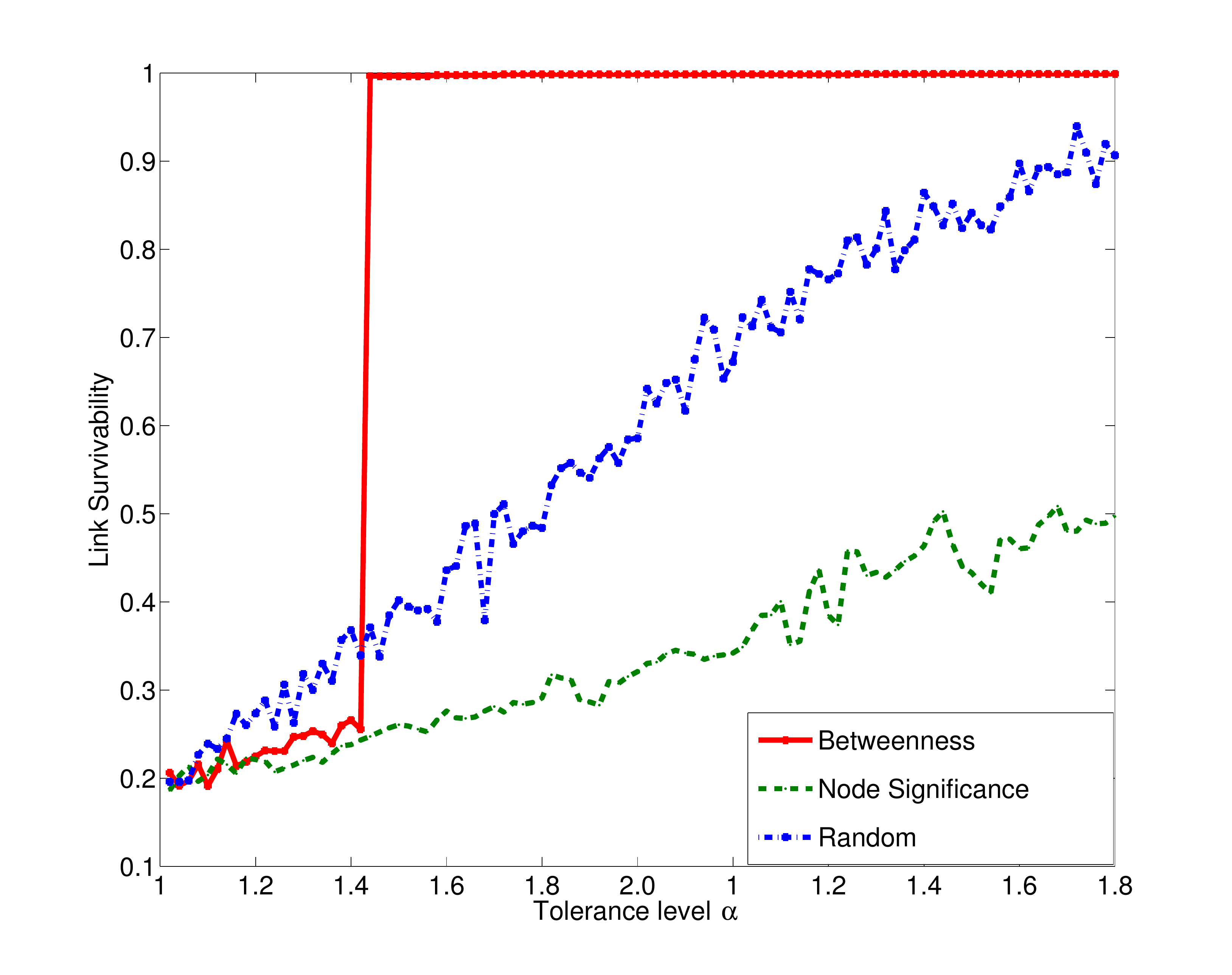}
	\label{fig:UCTELS}
	}
\caption{Link Survivability with respect to changes in Tolerance of the nodes in the network for different attack strategies.}
\label{fig:LinkSurv}
\end{center}
\vspace{-0.7cm}
\end{figure*}

\begin{figure*}[!htb]
\begin{center}	
	\subfloat[IEEE 118]{
	\includegraphics[width=.35\textwidth]{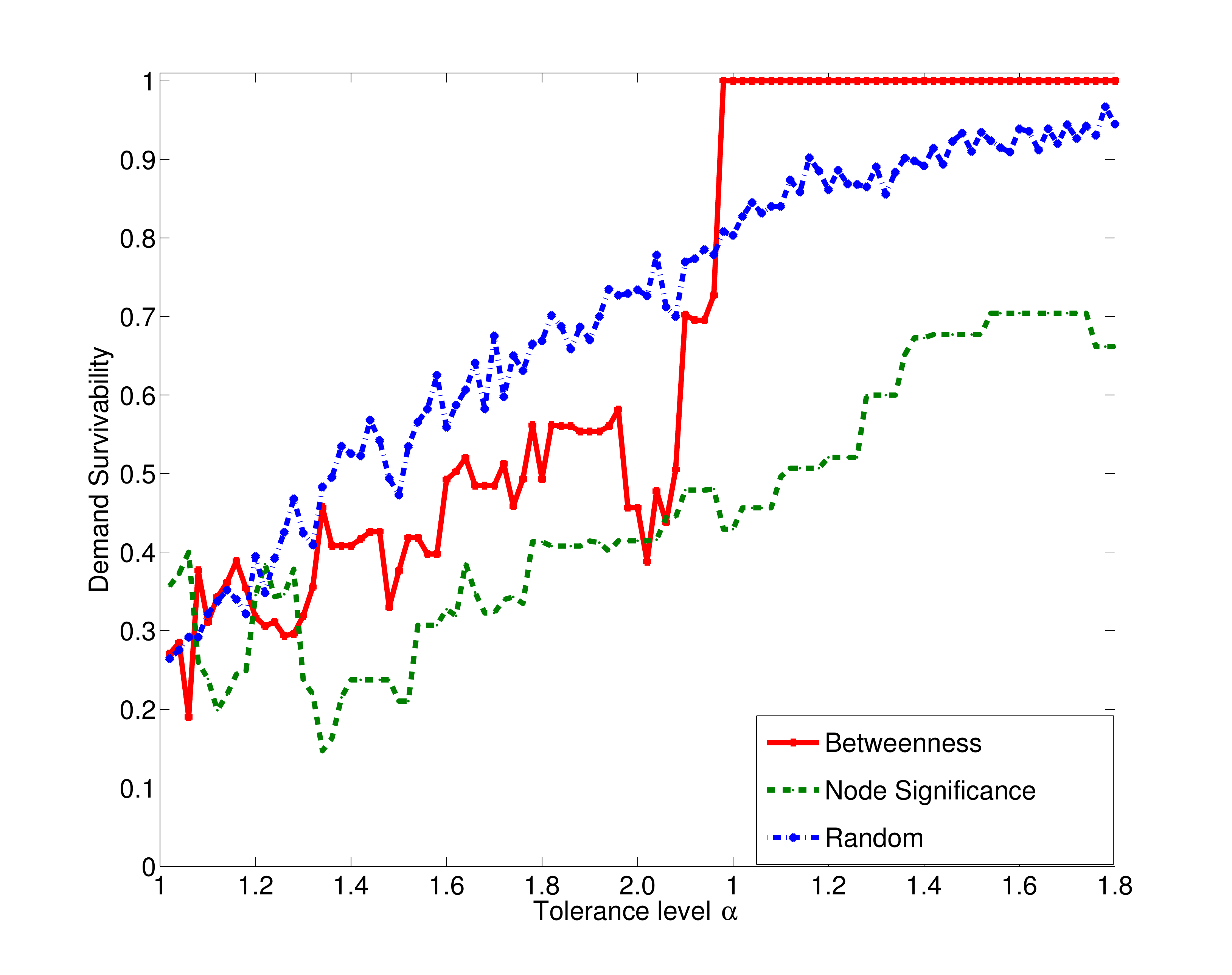}
	\label{fig:IEEEDS}
	}
	\subfloat[UCTE]{
	\includegraphics[width=.35\textwidth]{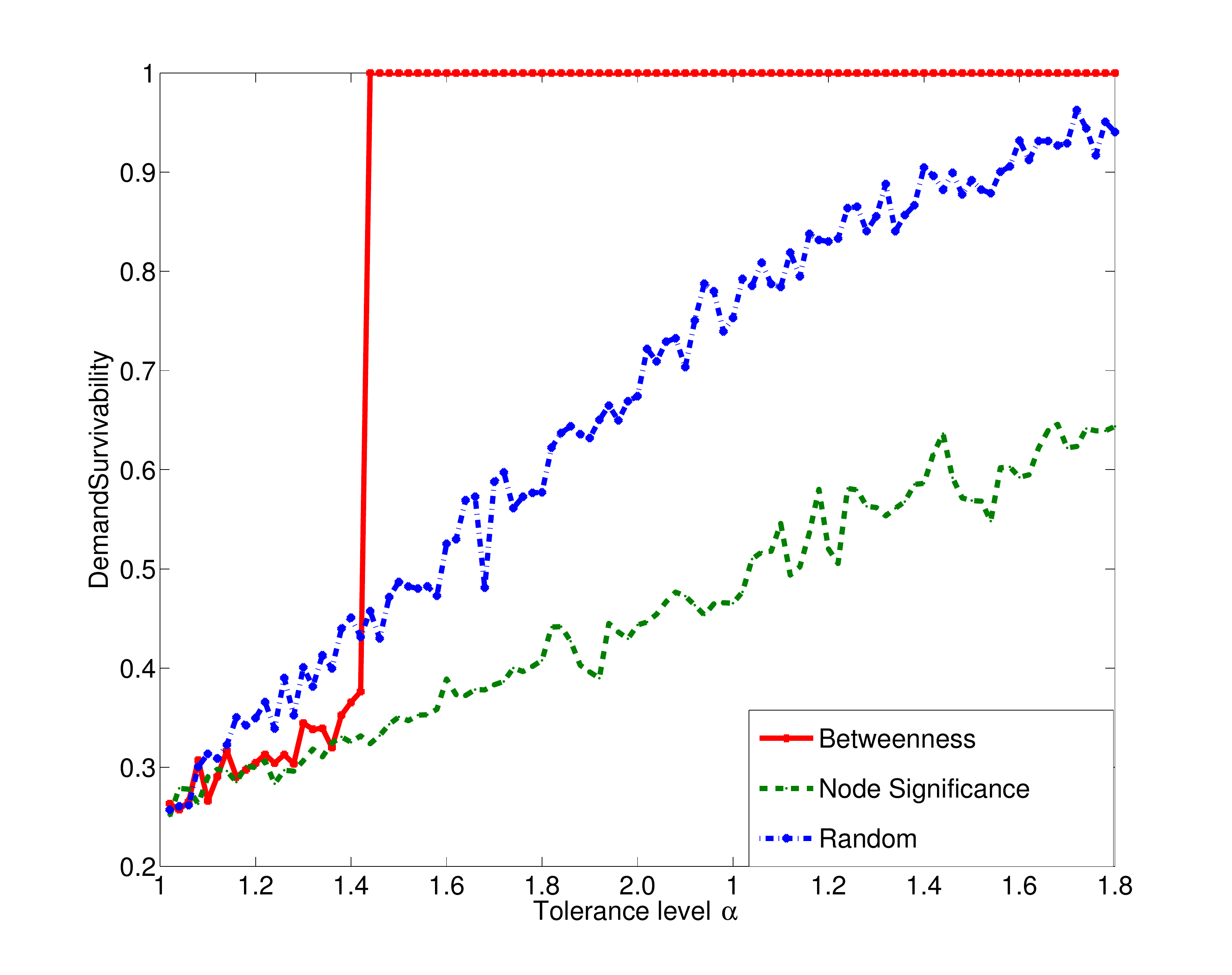}
	\label{fig:UCTEDS}
	}
\caption{Demand Survivability with respect to changes in Tolerance of the nodes in the network for different attack strategies.}
\label{fig:DemandSurv}
\end{center}
\vspace{-0.7cm}
\end{figure*}

This subsection considers the IEEE 118 system and the UCTE transmission grid with a summer profile to investigate their robustness under different attack strategies. Three different attack strategies are considered: (i) random attack, (ii) betweenness based attack, and (iii) node significance based attack. In each test grid a cascading failure is initiated based on these attack strategies, and simulated for different tolerance levels of the components. When the cascade subsides, the survivability (i.e. robustness) of a grids is quantified in terms of the fraction of energized links and satisfied power demand. Fig.~\ref{fig:LinkSurv} and ~\ref{fig:DemandSurv} show the result.  

For a random attack scenario (see Figs.~\ref{fig:LinkSurv} and ~\ref{fig:DemandSurv}) IEEE 118 and UCTE grids have a comparable level of robustness. For both grids, the relationship between the tolerance level $\alpha$ and the survivability of the grids is linear. As the tolerance level of components in a grid increases the survivability increases as well because the links are targeted randomly. Like scale-free network, power grids are significantly robust against random failures.

%


When attacking the networks based on betweenness centrality, a transition occurs at a threshold value of the tolerance parameter $\alpha$, below which the network collapses for any attack. In the UCTE network, the transition occurs at a tolerance parameter value of 1.22 (i.e. a loading level of 82\%), while in the IEEE 118 test systems it occurs at a tolerance parameter of 1.58 (i.e. a loading level of 63\%) suggesting that the UCTE summer configuration is more robust compared to the IEEE 118 test system configuration.   




In a node significance based attack scenario, the damage is much larger for both networks. Figs.~\ref{fig:LinkSurv} and ~\ref{fig:DemandSurv} show that even at high values of $\alpha$ the damage is significantly larger compared to a random failure or a betweenness based failure. The relationship between survivability and tolerance is not strictly linear but a general long term trend is visible suggesting that an increase in the tolerance parameter of the components in a power grid, results in a reduced damage. 


The simulation results on the robustness levels of different networks under different attack strategies show that the purely topological measures (e.g. betweenness centrality) underestimate the vulnerability of a power grid. A node significance based attack highlights much more damage caused to a power grid than theoretical centrality measures and random removal. This result is in line with results from Verma et al.~\cite{Verma13} that assess the effectiveness of different attack strategies.

\section{Conclusion and discussion}
\label{sec_Conclusion}

This paper introduces MATCASC, an open source, non-commercial MATLAB package to simulate cascading failures in power grids. MATCASC captures important characteristics of cascading failures including the line overloads and islanding effect by deploying a DC load flow analysis. It possesses mechanisms to deal with: (i) estimation of line power flow, (ii) estimation of the line capacities, (iii) offering different line removal options (e.g. a random removal or an attack-based removal), (iv) simulation of cascading line overloads outages, and (v) quantification of the damage due to the cascading failures. The applicability of MATCASC is demonstrated by (i) assessing the robustness level of IEEE test systems together with the real world UCTE network, and (ii) assessing the effectiveness of different attacks strategies on power grids.

MATCASC is mainly intended for the academic and research purposes. It is designed to be easily (i) understandable and (ii) extendible. The former feature of MATCASC enables the researchers to understand and use the tool easily in their cascading failure analysis study. The latter feature offers enough flexibility to extend it so that, ultimately, the tool evolves to an open source comprehensive cascading failure analysis tool capturing different cascading failures aspects. 

The future work will focus on incorporating other important aspects of cascading failures such as AC load flow analysis, the instability of frequency and voltage, and human/operator behaviour.   

%
%
%
 \subsection*{{\bf Acknowledgements}}
\noindent Partially funded by the NWO project \emph{RobuSmart}, grant number 647.000.001 and the KIC EIT project \emph{SES 11814 European Virtual Smart Grid Labs}.

\bibliography{RobustnessWS}

\begin{thebibliography}{10}
\providecommand{\url}[1]{#1}
\csname url@samestyle\endcsname
\providecommand{\newblock}{\relax}
\providecommand{\bibinfo}[2]{#2}
\providecommand{\BIBentrySTDinterwordspacing}{\spaceskip=0pt\relax}
\providecommand{\BIBentryALTinterwordstretchfactor}{4}
\providecommand{\BIBentryALTinterwordspacing}{\spaceskip=\fontdimen2\font plus
\BIBentryALTinterwordstretchfactor\fontdimen3\font minus
  \fontdimen4\font\relax}
\providecommand{\BIBforeignlanguage}[2]{{%
\expandafter\ifx\csname l@#1\endcsname\relax
\typeout{** WARNING: IEEEtran.bst: No hyphenation pattern has been}%
\typeout{** loaded for the language `#1'. Using the pattern for}%
\typeout{** the default language instead.}%
\else
\language=\csname l@#1\endcsname
\fi
#2}}
\providecommand{\BIBdecl}{\relax}
\BIBdecl

\bibitem{BlackoutRefBrazil}
J.~Conti, ``The day the samba stopped,'' \emph{Engineering Technology}, vol.~5,
  no.~4, 2010.

\bibitem{blackoutReport}
``{U.S.- Canada Power System Outage Task Force, Final Report on the August 14th
  Blackout in the United States and Canada: Causes and Recommendations},''
  2004.

\bibitem{Baldick2008}
{R. Baldick et al.}, ``Initial review of methods for cascading failure analysis
  in electric power transmission systems,'' in \emph{IEEE PES General Meeting},
  2008.

\bibitem{Eeten2011}
M.~Van~Eeten, A.~Nieuwenhuijs, E.~Luiijf, M.~Klaver, and E.~Cruz, ``The state
  and the threat of cascading failure across critical infrastructures,''
  \emph{Public Administration}, vol.~89, no.~2, 2011.

\bibitem{Schneider2013}
C.~M. Schneider, N.~Yazdani, N.~Araujo, S.~Havlin, and H.~J. Herrmann,
  ``Towards designing robustn coupled networks,'' \emph{Scientific Reports},
  vol.~3, 2013.

\bibitem{Buldyrev2010}
S.~V. Buldyrev, R.~Parshani, G.~Paul, H.~E. Stanley, and S.~Havlin,
  ``{Catastrophic cascade of failures in interdependent networks},''
  \emph{Nature}, vol. 464, no. 7291, Apr. 2010.

\bibitem{Vaiman2012}
{M. Vaiman et al.}, ``Risk assessment of cascading outages: Methodologies and
  challenges,'' \emph{IEEE Transactions on Power Systems}, vol.~27, no.~2,
  2012.

\bibitem{Kirschen2004}
D.~S. Kirschen, D.~Jayaweera, D.~P. Nedic, and R.~N. Allan, ``A probabilistic
  indicator of system stress,'' \emph{IEEE Transactions on Power Systems},
  vol.~19, 2004.

\bibitem{Hardiman2004}
R.~Hardiman, M.~Kumbale, and Y.~Makarov, ``An advanced tool for analyzing
  multiple cascading failures,'' in \emph{Int. Conf. on Prob. Methods Applied
  to Power Systems}, 2004.

\bibitem{Carreras2004}
B.~A. Carreras, V.~E. Lynch, I.~Dobson, and D.~E. Newman, ``Complex dynamics of
  blackouts in power transmission systems,'' \emph{Chaos}, vol.~14, 2004.

\bibitem{Chen2005}
J.~Chen, J.~S. Thorp, and I.~Dobson, ``Cascading dynamics and mitigation
  assessment in power system disturbances via a hidden failure model,''
  \emph{International Journal of Electrical Power Energy Systems}, vol.~27,
  no.~4, 2005.

\bibitem{Liao2004}
H.~Liao, J.~Apt, and S.~Talukdar, ``Phase transitions in the probability of
  cascading failures,'' in \emph{Electricity Transmission in deregulated
  markets}, 2004.

\bibitem{Grainger1994}
J.~J. Grainger, J.~Stevenson, and D.~William, \emph{Power System
  Analysis}.\hskip 1em plus 0.5em minus 0.4em\relax McGraw-Hill, 1994.

\bibitem{Hertem2006}
{D. Van Hertem et. al}, ``Usefulness of dc power flow for active power flow
  analysis with flow controlling devices,'' in \emph{IEEE Conf. on AC and DC
  Power Transmission}, 2006.

\bibitem{Bao2009}
Z.~J. Bao, Y.~J. Cao, G.~Z. Wang, and L.~J. Ding, ``Analysis of cascading
  failure in electric grid based on power flow entropy,'' \emph{Physics Letters
  A}, vol. 373, 2009.

\bibitem{Glover2001}
J.~D.~D. Glover and M.~S. Sarma, \emph{Power System Analysis and Design},
  3rd~ed.\hskip 1em plus 0.5em minus 0.4em\relax Brooks/Cole Publishing Co.,
  2001.

\bibitem{Motter2002}
A.~E. Motter and Y.-C. Lai, ``Cascade-based attacks on complex networks,''
  \emph{Phys Rev E}, vol.~66, 2002.

\bibitem{Pepyne2007}
D.~Pepyne, ``Topology and cascading line outages in power grids,''
  \emph{Journal of Systems Science and Systems Engineering}, vol.~16, 2007.

\bibitem{Zimmerman2011}
R.~Zimmerman, C.~Murillo-Sanchez, and R.~Thomas, ``Matpower: Steady-state
  operations, planning, and analysis tools for power systems research and
  education,'' \emph{Power Systems, IEEE Transactions on}, vol.~26, no.~1,
  2011.

\bibitem{Mieghem2006}
P.~Van~Mieghem, \emph{Performance analysis of communications networks and
  systems}.\hskip 1em plus 0.5em minus 0.4em\relax Cambridge University Press,
  2006.

\bibitem{Koc2013}
Y.~Ko\c{c}, M.~Warnier, R.~E. Kooij, and F.~Brazier, ``A robustness metric for
  cascading failures by targeted attacks in power networks,'' in
  \emph{Proceedings of the IEEE Conference on Networking Sensing and Control},
  2013.

\bibitem{Koc2013_2}
Y.~Ko\c{c}, M.~Warnier, R.~E. Kooij, and F.~M. Brazier, ``An entropy-based
  metric to quantify the robustness of power grids against cascading
  failures,'' \emph{Safety Science}, vol.~59, 2013.

\bibitem{Verma13}
T.~Verma, W.~Ellens, and R.~Kooij, ``Context-independent centrality measures
  underestimate the vulnerability of power grids,'' \emph{{Int. Journal of
  Critical Infrastructures}}, 2013.

\end{thebibliography}
\bibliographystyle{IEEEtran}

\end{document}